\documentclass[preprint,aps,floatfix,superscriptaddress,pre]{revtex4}

\usepackage{amsmath, amsthm, amssymb, graphicx}
\usepackage{color}
\usepackage{multirow}
\usepackage{ulem}
\usepackage{float}
\usepackage{bm}
\usepackage{subfiles}
\usepackage{siunitx}
\usepackage{booktabs}
\usepackage{array}

\begin{document}

\title{Machine learning generated configurations in presence of a conserved quantity: a cautionary tale}
\author{Ahmadreza Azizi}
\affiliation{Department of Physics, Virginia Tech, Blacksburg, VA 24061-0435, USA}
\affiliation{Center for Soft Matter and Biological Physics, Virginia Tech, Blacksburg, VA 24061-0435, USA}
\author{Michel Pleimling}
\affiliation{Department of Physics, Virginia Tech, Blacksburg, VA 24061-0435, USA}
\affiliation{Center for Soft Matter and Biological Physics, Virginia Tech, Blacksburg, VA 24061-0435, USA}
\affiliation{Academy of Integrated Science, Virginia Tech, Blacksburg, VA 24061-0563, USA}
\date{\today}

\begin{abstract}
We investigate the performance of machine learning algorithms trained exclusively with configurations obtained from importance sampling 
Monte Carlo simulations of the two-dimensional Ising model with conserved magnetization. For supervised machine learning,
we use convolutional neural networks and find that the corresponding output not only allows to locate the phase transition point with
high precision, it also displays a finite-size scaling characterized by an Ising critical exponent. For unsupervised learning, restricted Boltzmann machines (RBM) are
trained to generate new configurations that are then used to compute various quantities. We find that RBM is
incapable of recognizing the conserved quantity and generates configurations with magnetizations and energies forbidden in the
original physical system. The RBM generated configurations result in energy density probability distributions with incorrect weights as
well as in wrong spatial correlations. We show that shortcomings are also encountered when training RBM with configurations obtained
from the non-conserved Ising model.
\end{abstract}

\maketitle

\section{Introduction}
In recent years machine learning applications have seen a rapid proliferation in almost all fields of science. In condensed matter physics
machine learning models, that aim at learning the probability distribution that the input data have been sampled from,
have been used to investigate phase transitions \cite{Wang16,Wetzel17,Wang17,Hu17,Wang18,Ponte17,Zhang19a,Alexandrou17,Torlai16,Morningstar18,Rao18,
Carrasquilla17,Nieuwenburg17,Kim18,Zhang19,Shiina20,Tanaka17,Li18,Beach18}, 
characterize quantum circuits \cite{Wiebe14,Kieferova17,Benedetti17,Benedetti18}, predict crystal structures \cite{Curtarolo03,Morgan05,Fischer06},
and learn renormalization group flow \cite{Li18a,Koch18}, to name but of few areas of applications.

Machine learning methods, which encompass methods as
diverse as principal component analysis \cite{Wang16,Wetzel17,Wang17,Hu17,Wang18}, support vector machines \cite{Ponte17,Zhang19a}, 
and variational autoencoders \cite{Wetzel17,Alexandrou17}, in addition to various neural network architectures 
\cite{Zhang19a,Carrasquilla17,Nieuwenburg17,Kim18,Zhang19,Shiina20,Tanaka17,Li18,Beach18}, have been applied to study various properties
of physical systems. These models can be grouped into the two broad categories of supervised and unsupervised learning.
Supervised learning methods have successfully identified phases of matter and located phase transition points.
In many of these approaches the algorithms are trained over physical configurations, often obtained from importance sampling Monte Carlo
simulations, after which they are used to predict the class of a test configuration. For a system with an order-disorder transition, configurations
are typically classified as ordered or as disordered. The output from these classifiers, that display a behavior equivalent to
an order parameter \cite{Carrasquilla17}, can reliably determine the location of phase transition points. For systems with no clear order
parameter, machine learning output has been shown to play a role similar to an order parameter which can then be exploited to
locate a phase transition \cite{Wang16,Wetzel17,Tanaka17,Wang17,Hu17,Zhang19}. For example, in \cite{Wang16} principle component analysis
was used to extract from spin configurations of the conserved-magnetization Ising model the structure factor which allowed to locate the
critical point.

The physics community has also paid much attention to generative learning models,
a subset of unsupervised learning methods, that reconstruct configurations 
for classical \cite{Wetzel17,Carrasquilla17,Torlai16} or quantum \cite{Carleo17,Chng17,Broecker17,Torlai18} systems.
The idea behind generative models is to learn the hidden probability distribution underlying the unlabeled input data. Generative
adversarial networks \cite{Goodfellow14} and variational autoencoders \cite{Kingma13} are two types of generative models
used for classical systems \cite{Wetzel17,Liu17a,Cristoforetti17}. However, more broadly used are restricted Boltzmann machines (RBM)
\cite{Torlai16,Morningstar18,Huang17,Nomura17,Pilati19,Melko19,Yu19,Vieijra20}.

Machine learning studies of classical spin systems have almost exclusively focused on non-conserved models. However, conserved quantities
can have a major impact on the physical properties of a system as they put strong constraints on the accessible parts of configuration 
space. In \cite{Wang16} and \cite{Ponte17} learning algorithms, based either on principal component analysis (PCA) or support vector machines (SVM),
were applied on configurations obtained from Monte Carlo (MC) simulations of the two-dimensional conserved-magnetization Ising model. It was shown 
that for zero magnetization configurations (i.e. configurations with exactly the same number of up and down spins) the output of the
algorithms behaved like an order parameter which then allowed to locate the phase transition point. 

In the following we report the results of discriminative and generative machine learning on
training configurations obtained from Monte Carlo simulations of the Ising model with Kawasaki dynamics 
and constant magnetization \cite{Kawasaki66}.
As binary classifier we use a convolutional neural network (CNN) \cite{Lawrence97} and show that the CNN output not only behaves like
an order parameter and allows to locate with high precision the critical point separating the ordered and disordered phases, it also
displays a finite-size scaling governed by the critical exponent $\nu$. The outputs of PCA and SVM from datasets obtained
for the constant-magnetization Ising model do not exhibit the same scaling
property. For the generative machine learning we use a restricted Boltzmann machine, trained on datasets obtained from the Ising model with
constant (not necessarily zero) magnetization. While RBM allows to obtain good estimates for the average energy, a closer look reveals
major shortcomings with the RBM generated configurations. Indeed, RBM is not capable of identifying a conserved quantity and therefore generates
configurations with magnetization and energy values forbidden in the conserved-magnetization Ising model. We also observe systematic differences 
between RBM generated configurations and MC generated configurations in the weights of the energy density probability distribution function.
Finally, spatial correlations computed using the RBM configurations deviate systematically from spatial configurations obtained in Monte Carlo
simulations with spin exchanges. 
We also revisit machine learning for the non-conserved Ising model and find the same issues with the energy density probability distribution
and with the spatial correlations.

\section{Two-dimensional Ising model with conserved magnetization}

As a simple model with a conserved quantity we consider in this work the Ising model with conserved magnetization.
Every point with coordinates $(i,j)$, $i,j=1, \cdots, L$, on a square lattice of linear length $L$ is characterized by a classical 
variable $S_{i,j}$ that can take on only the two values 1 and $-1$.
Setting the coupling constant equal to 1, the energy of an arrangement of these spin variables is given by
\begin{equation}
{\cal H} = - \sum_{i,j=1}^L \left( S_{i,j} S_{i+1,j} + S_{i,j} S_{i,j+1} \right)
\end{equation}
with the periodic boundary conditions $S_{L+1,j} = S_{1,j}$ and $S_{i,L+1}=S_{i,1}$. The magnetization density is 
\begin{equation}
M = \frac{1}{L^2} \sum_{i,j=1}^L S_{i,j} ~,
\end{equation} 
whereas the energy density is
\begin{equation}
E = \frac{1}{L^2} \sum_{i,j=1}^L S_{i,j} \left( S_{i+1,j} + S_{i,j+1}  \right) ~.
\label{E}
\end{equation}
In the thermodynamic limit and without additional constraints the
two-dimensional Ising model undergoes at the critical temperature (setting $k_B=1$) $T_c=2/\ln(1+ \sqrt{2})$ a continuous phase transition separating
the ordered, magnetized phase with non-vanishing magnetization from the disordered phase with zero magnetization. 

Fixing the magnetization at some value $M_0$ narrows the space of possible configurations. For $M=M_0=0$ only configurations
with exactly 50\% of the spins taking on each of the two possible values are accessible. Obviously, the magnetization then does not
display anymore the typical behavior of an order parameter when crossing the critical temperature.

We generate for a fixed magnetization density $M = M_0$ independent configurations at a temperature $T$ through standard importance sampling Monte 
Carlo simulations with spin exchange. These configurations are then used for two purposes: (1) to train the machine learning
algorithms and (2) to compute energy density probability distributions as well as thermal averages of 
the energy density $\varepsilon = \left< E \right>$ (here and in the following $\left< \cdots \right>$
indicates an average over configurations), the absolute value of the magnetization density
\begin{equation}
| M | = \left< \frac{1}{L^2} | \sum_{i,j=1}^L S_{i,j} | \right>~,
\end{equation}
and the space-dependent correlations
\begin{equation}
C(r) = \left< \frac{1}{2 L^2} \sum_{i,j=1}^L S_{i,j} \left( S_{i+r,j} + S_{i,j+r}  \right) \right>~
\label{C}
\end{equation}
and compare the values of these quantities with those obtained from the configurations created through machine learning.

\begin{figure}
 \centering \includegraphics[width=0.20\columnwidth,clip=true]{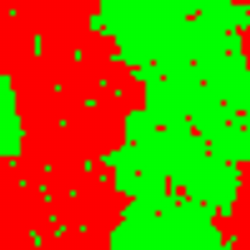} ~~ \includegraphics[width=0.20\columnwidth,clip=true]{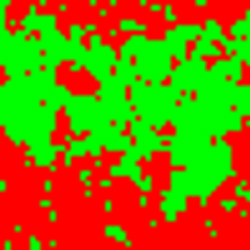}
~~ \includegraphics[width=0.20\columnwidth,clip=true]{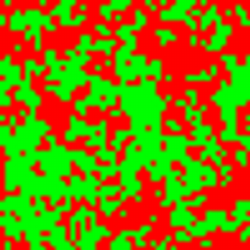}
 \caption{Typical configurations for a system with $L=50$, $M_0=0$ and (from left to right) $T=2$, $T=T_c$, and $T=3$.
}
\label{fig1}
\end{figure}

Typical configurations with zero magnetization are shown in Fig. \ref{fig1}. Below $T_c$ the system phase separates into two parts with different majority spins.
The configuration with the lowest energy, which is the dominating configuration at very low temperatures, is the one where the system is separated
into two perfectly ordered halves, with straight interfaces separating the two parts. Increasing the temperature, the interfaces roughen and
the halves are less well ordered, see the configuration at $T=2$. Above $T_c$ the large compact regions are broken up and complicated, interlocked
clusters remain.

\section{Convolutional Neural Network: phase transition with conserved order parameter}

In classical physics, machine learning, both supervised and unsupervised, has quickly become a much used tool for the 
study of phase transitions, as witnessed by recent investigations of spin systems
using techniques as diverse as principal component analysis \cite{Wang16,Wetzel17,Wang17,Hu17,Wang18}, support vector machines \cite{Ponte17,Zhang19a}, 
variational autoencoders \cite{Wetzel17,Alexandrou17}, 
Boltzmann machines \cite{Torlai16,Morningstar18,Rao18}, fully connected neural networks \cite{Carrasquilla17,Nieuwenburg17,Kim18,Zhang19,Shiina20} as well as 
convolutional neural networks \cite{Carrasquilla17,Tanaka17,Li18,Beach18,Zhang19,Zhang19a}. The vast majority of these studies focused on systems without
conserved quantities. Only two studies, one using principal component analysis (PCA) \cite{Wang16}, the other support vector machines (SVM) \cite{Ponte17},
briefly discussed the two-dimensional Ising model with conserved order parameter. Our goal in the following is to use a convolutional neural network (CNN)
based classifier and investigate the phase transition in the two-dimensional Ising model using exclusively configurations with zero magnetization.
As we will show, for the two-dimensional Ising model with zero magnetization the output of the CNN displays a critical finite-size scaling behavior not 
seen when using PCA or SVM. 

Our model is a binary classifier with convolutional layers that determines whether a test
configuration is in the ordered phase or not. CNNs, which are considered to be the most successful models in image processing problems \cite{Lawrence97},
are able to extract important features of images, i.e, they learn boundaries of different objects in a
given image and classify them accordingly. This is an important property when learning the phase transition in
systems with conserved order parameter. Indeed, for $M_0=0$ there is no majority spin state, and configurations are composed of similar clusters
for all spin states. Therefore, unlike fully connected neural network models that merely use the number of majority spins in
configurations to learn the different thermodynamic phases and therefore fail in cases with conserved magnetization \cite{Carrasquilla17}, convolutional neural layers are well suited 
for the system at hand as they learn the differences between the shapes of spin clusters in the different (ordered and disordered) phases.

\begin{figure}
 \centering \includegraphics[width=0.75\columnwidth,clip=true]{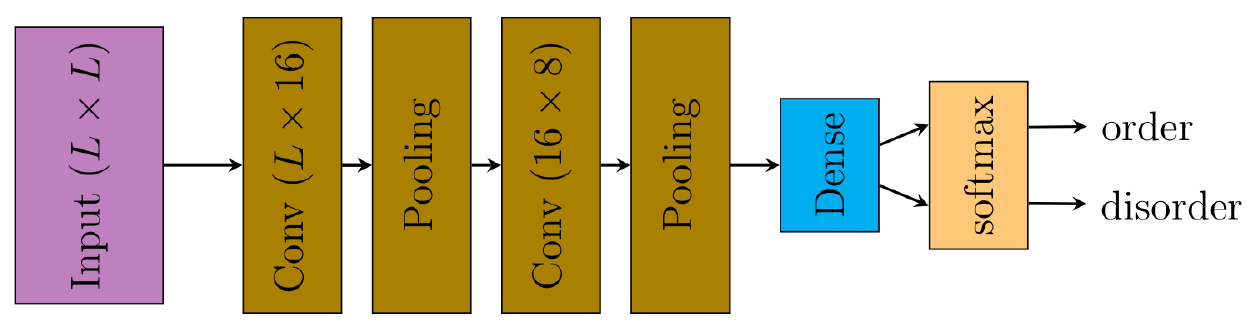}
 \caption{Structure of the convolutional neural network used to investigate the phase transition in the two-dimensional Ising model with conserved magnetization.
$L$ is the linear dimension of the square lattice.
}
\label{fig2}
\end{figure}

Figure \ref{fig2} depicts the structure of our CNN model. It includes two sets of convolutional layer
with pooling layer that are followed by a dense layer with a softmax activation function.
Therefore the CNN model output is a real number in the range [0, 1] which is the probability
of the given configuration being in the ordered phase (of course, as the sum of the probabilities to be in the ordered or disordered
phase is one, the output also provides us with the probability that the given configuration is in the disordered phase).

\begin{figure}
 \centering \includegraphics[width=0.60\columnwidth,clip=true]{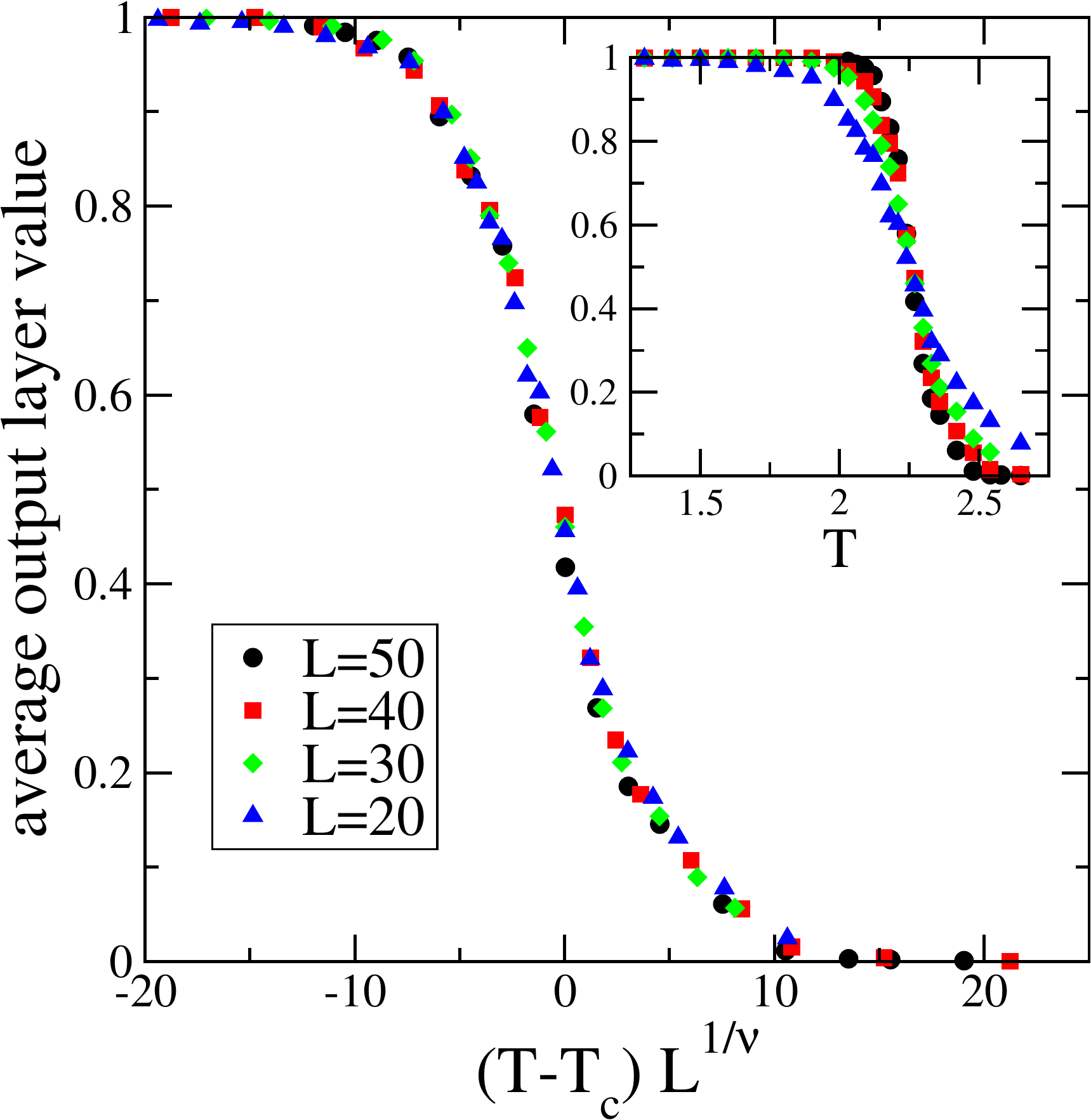}
 \caption{Average output layer value over the test configurations for different system sizes as function of temperature. As shown in the inset, this quantity displays the
expected behavior of an order parameter close to a phase transition. Zooming in into the temperature region close to $T_c \approx 2.269$ reveals that the data for all the
different system sizes reach the value 0.5 in very close vicinity to $T_c$. As shown in the main figure the average output layer value displays a finite-size scaling behavior governed
by the Ising exponent $\nu = 1$. Error bars are comparable to the size of the symbols.
}
\label{fig3}
\end{figure}

Following the standard protocol for supervised learning of phases, we generate a large number of configurations from importance sampling Monte Carlo simulations
of square Ising systems of linear dimension $L$ with zero magnetization and Kawasaki spin exchange dynamics. For every system size $L=20, 30, 40, 50$ we generate $10^5$ configurations
at temperatures $T=1.0+(n-1) \Delta T$ with $\Delta T = 0.1$ and $n=1, \cdots, 23$. These configurations are then split into training sets (70\% of configurations generated at
each of the 23 temperatures), validation sets (10\% of generated configurations) and test sets (20\% of generated configurations). For the training, configurations generated at 
temperatures $T < T_c$ receive the label '1', whereas configurations generated at $T > T_c$ are labeled as '0'. Convergence of the training process, during which the
learning algorithm is optimized for a number of epochs using the entire training set, is evaluated by the validation dataset. After each training epoch
the model accuracy is measured and the training process is stopped when for three consecutive epochs the model accuracy on the validation dataset is not improved.
Finally, the test dataset is used to evaluate the model's performance.

Fig. \ref{fig3} summarizes our results for the average output layer value over the test configurations. As shown in the inset, this quantity, similarly to the outputs
obtained from PCA \cite{Wang16} and SVM \cite{Ponte17}, displays the generic behavior of the order parameter in a system with an order-disorder phase transition, with
a value close to 1 at low temperatures (ordered phase) and a value close to zero at high temperatures (disordered phase). The deviations from the value 1 below $T_c$ and
from the value 0 above $T_c$ are the results of false classifications. These misclassifications have a physical origin and reflect the increase of
fluctuations and of diversity of configurations close to a critical point. For the different system sizes, this quantity
takes on the value 0.5 in the temperature interval $\left[ 2.25, 2.27 \right]$, in 
very close proximity to the known critical temperature $T_c = 2/\ln(1+ \sqrt{2}) \approx 2.269$. The average output layer value displays 
a critical finite-size scaling behavior, see the main panel in Fig. \ref{fig3}, as it is a function of $\left( T - T_c \right) L^{1/\nu}$ with the Ising exponent $\nu =1$. 
This critical finite-size scaling, which has not been observed in the earlier investigations of the
conserved-order-parameter Ising model using machine learning techniques
\cite{Wang16,Ponte17}, is similar to that encountered in supervised learning of non-conserved spin systems close to their critical point \cite{Carrasquilla17,Shiina20}.

\section{Restricted Boltzmann Machine: space-dependent correlations}
Generative learning, with the goal to capture the probability distribution function underlying the input data and produce new data similar to the input data,
is a demanding task. In order to  perform this task Torlai and Melko \cite{Torlai16} have used a restricted Boltzmann machine (RBM), a stochastic neural network \cite{Ackley85,Hinton10}, 
and have shown that the spin configurations generated in that way yield for the non-conserved Ising model reasonable values for quantities like the average
magnetization and energy densities. Subsequently RBMs have been used successfully on other classical \cite{Rao18} and, especially, quantum systems
\cite{Amin18,Carleo17,Deng17}. Similar generative
approaches have also been used for neural network renormalization group studies \cite{Li18a,Koch18},
as well as in proposals to exploit machine learning for accelerating Monte Carlo simulations \cite{Huang17,Liu17}. 

What has been missing in earlier studies using RBMs for the investigation of classical spin systems is a stringent test of the quality of 
the probability distribution underlying the process of creating representative configurations that are then used to compute (thermal) averages. As already mentioned,
RBM generated configurations yield averages for magnetization density and energy density that agree well with those obtained from importance sampling
Monte Carlo simulations \cite{Torlai16}, but then even rather dissimilar probability distributions can result for some quantities in averages that roughly agree. In the following
we show results that point to major differences between RBM generated configurations and configurations obtained from Monte Carlo simulations
and used to train the RBM. Most revealing
will be the comparison of the energy density probability distributions
as well as of spatial correlations (\ref{C}). Some of these differences come from the fact that RBM can not cope in good ways with
conserved quantities. However, we will show via the spatial correlations in the non-conserved Ising model that 
the observed deficiencies are more general and are not restricted to systems
that exhibit a conserved quantity.

\begin{figure}
 \centering \includegraphics[width=0.75\columnwidth,clip=true]{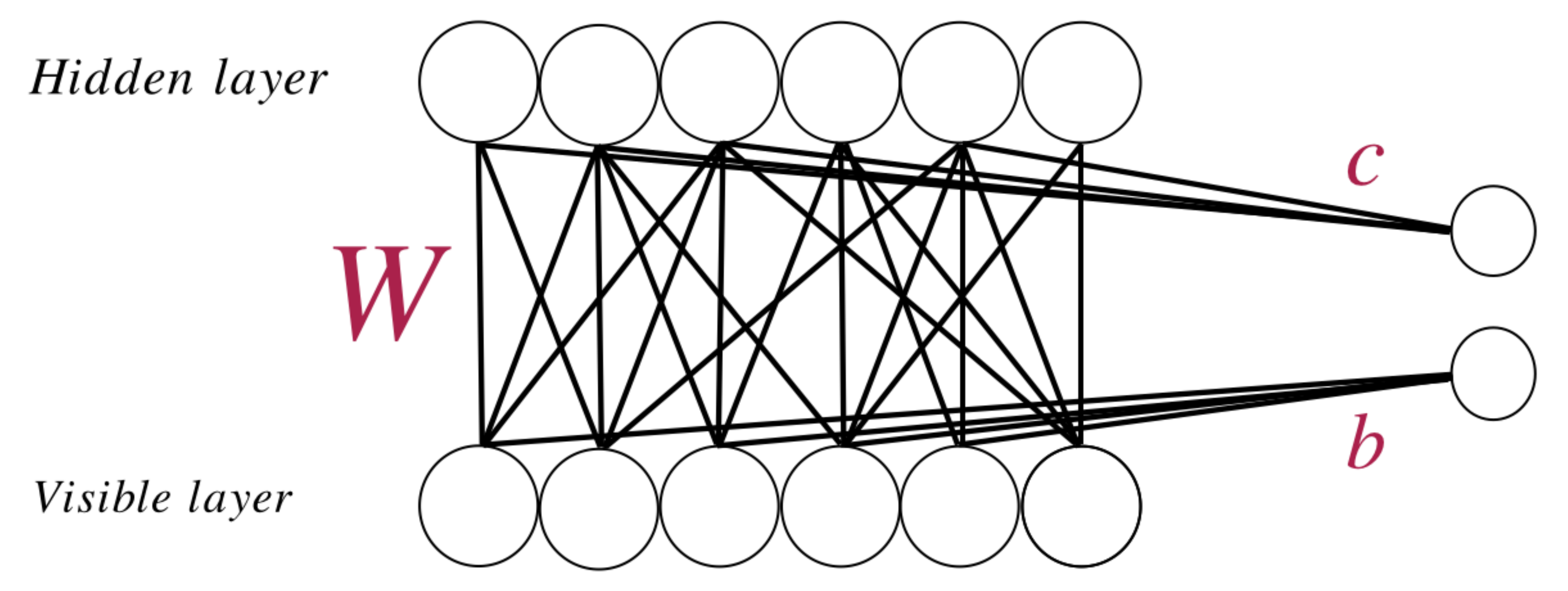}
 \caption{Sample architecture of RBM. In our implementation the number of nodes in the hidden layer is chosen to be identical to the number of nodes in 
the visible layer. The two nodes on the right hand side represent the bias terms in Eq. (\ref{E_RBM}).
}
\label{fig4}
\end{figure}

\subsection{Some technical details}
An earlier thorough investigation has shown that a shallow RBM is more efficient to learn the physical properties of Ising models than a deep 
generalization \cite{Morningstar18}. Taking this into account, we apply only one hidden layer in our RBM architecture, i.e. we have only two layers, the visible
layer $V$ and one hidden layer $H$, see Fig. \ref{fig4}. RBMs have been discussed in detail elsewhere \cite{Ackley85,Hinton10,Torlai16,Morningstar18,Rao18}. We give in the following
only a brief description, with the main intend to provide values for the parameters used in our implementation.

Denoting the nodes in the input layer as $V=\left\{ v_1 , \cdots , v_n \right\}$ and the nodes in the hidden layer as $H = \left\{ h_1 , \cdots , h_m \right\}$, with the hidden
nodes only taking on binary values 0 and 1, the joint probability distribution is defined as
\begin{equation}
P(V,H) = \frac{1}{Z} e^{-E(V,H)}
\end{equation}
with the ``energy''
\begin{equation} \label{E_RBM}
E(V,H) = - V^T b - c^T H - V^T W H
\end{equation}
and the ``partition function'' $Z = \sum_{V,H} e^{-E(V,H)}$. Here, $b$, $c$, and $W$ contain the model parameters that are trainable through the optimization scheme. 
It was shown in \cite{Torlai16} and \cite{Morningstar18} that for Ising systems $m=n=L^2$ is an appropriate choice for the number of hidden nodes, and we will present in the
following results with this number of hidden nodes. We also explored cases with $m > n$, but this did not yield substantial improvements.

Summarizing the model parameters as $\theta = \left\{ b, c, W \right\}$ and performing the summation of the hidden layer nodes that take on only the values 0 and 1 yields
the likelihood function
\begin{equation}
P(V | \theta) = \frac{1}{Z} e^{V^T b} \prod_{j=1}^n \left( 1 + e^{c_j + V^T w_j} \right)
\end{equation}
where $w_j$ is the $j$th column in $W$. As optimization method we apply a gradient descent based scheme to update the parameters $\theta$ at each iteration:
\begin{equation}
\theta_{t+1} - \theta_t = \alpha ~ \nabla \log P(V,\theta)
\end{equation}
with the learning rate $\alpha$. The gradient operator acts on all vector and matrix elements in $\theta = \left\{ b, c, W \right\}$.
More specifically, we use the learning rate $\alpha = 5 \times 10^{-3}$, the `Adam' optimization scheme \cite{Kingma14}, and
the Contrastive Divergence algorithm CD-k with $k=10$ steps \cite{Hinton02}.
Typically 10,000 spin configurations obtained from Monte Carlo simulations of the Ising model are used for the training.

Once the training phase is done and the optimized values $b^*$, $c^*$, and $W^*$ are found for the parameters, Gibbs sampling is used to generate new configurations.
Like CD-k, Gibbs sampling performs a Markov chain between the visible and invisible layers, but this time the chain starts from a random initial configuration
in the visible layer. 
The values $H = \left\{ h_1, \cdots , h_n \right\} $ of the nodes in the invisible layer are
computed from $P(H | \theta^*, V)$. In a second step the values of the visible layer are updated with the help
of the probability distribution $P(V | \theta^*, H)$: the probability that the node $j$ in the visible layer takes on the value 1 is
\begin{equation}
P(v_j =1  | \theta^*, H) = \frac{\exp \left( b_j^* + h_i w_{ij}^* \right) }{1+ \exp \left( b_j^* + h_i w_{ij}^* \right) }~,
\end{equation}
where $h_i$ is the value of node $i$ in the hidden layer. After drawing a random number $p$ from the interval $\left( 0, 1 \right)$, the node $j$ in the visible layer
is updated to the value $v_j=1$ when $p \le P(v_j =1  | \theta^*, H)$ and to the value $v_j=-1$ otherwise. This is repeated $k$ times until convergence is reached.
We investigated a rather wide range of number of steps $k$, ranging from 2 to 50, and found that in most cases
$k=10$ is enough to reach the final converged state. It is worth mentioning that increasing the number of training configurations to 20,000
did not significantly improve the quality of the RBM configurations.

\begin{figure}
 \centering \includegraphics[width=0.60\columnwidth,clip=true]{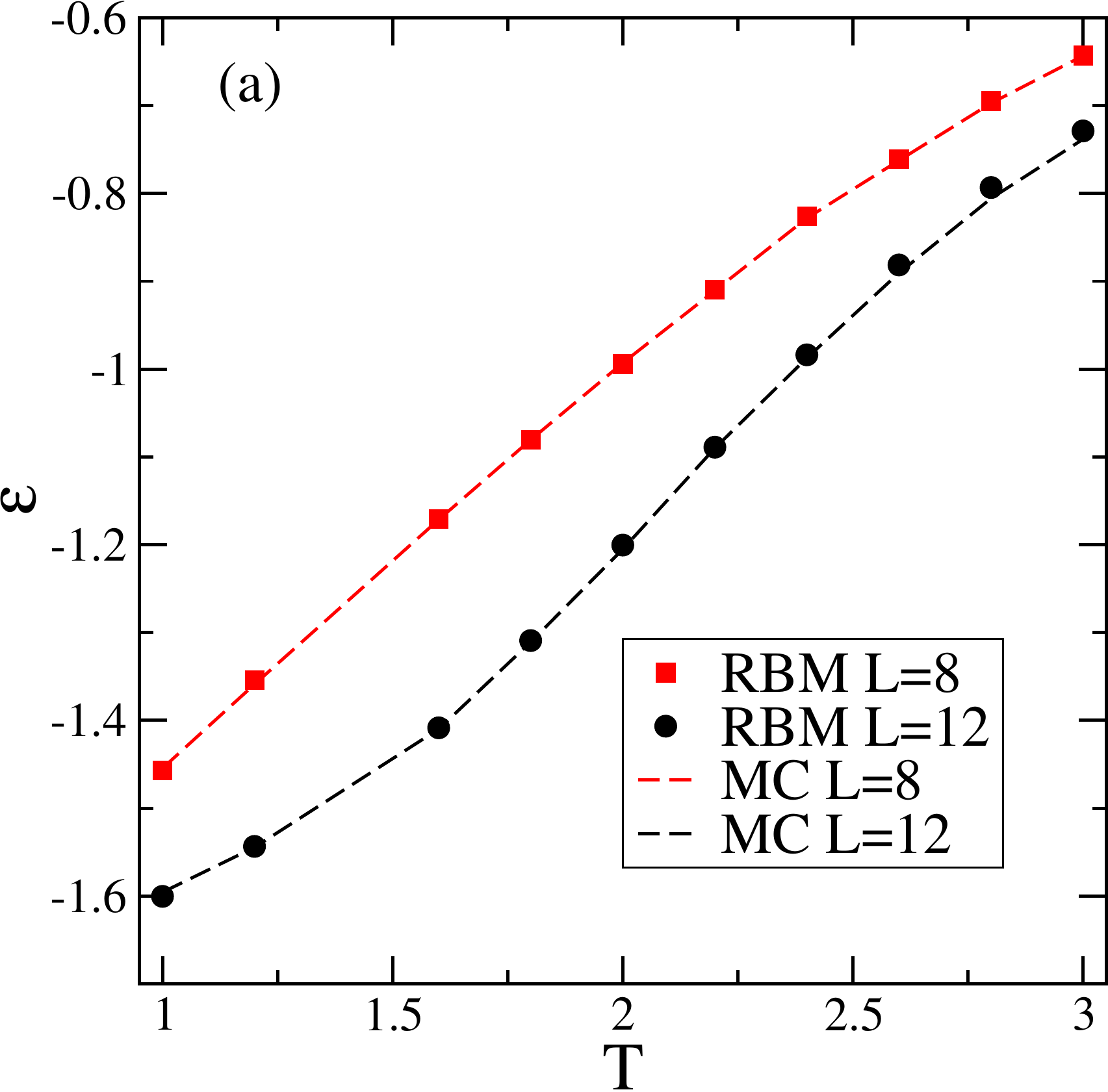}\\
 \centering \includegraphics[width=0.60\columnwidth,clip=true]{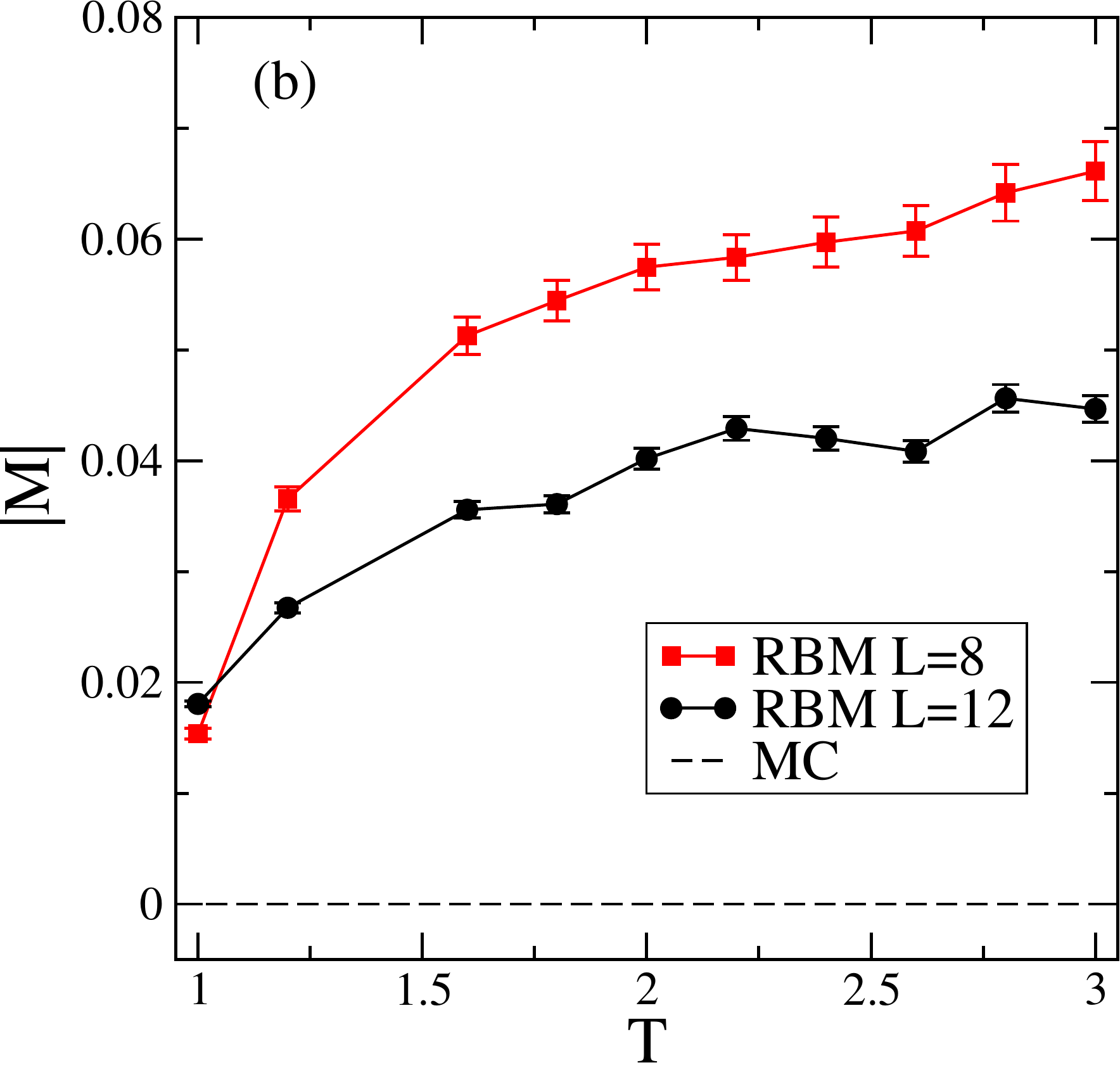}
 \caption{Comparison of averages obtained from RBM generated configurations (symbols) and from configurations generated in importance sampling Monte
Carlo (MC) simulations of the conserved-magnetization Ising model with $M_0=0$ (dashed lines). The RBM machine has been trained using the MC configurations. Data for two
different system sizes are displayed. Shown are (a) the energy density and (b) the magnetization density. 
}
\label{fig5}
\end{figure}

\subsection{Ising systems with conserved magnetization}
We start by discussing our results for the Ising model with conserved magnetization, which has been the main subject of our investigation. This is
followed by a brief discussion of the spatial correlations in the non-conserved Ising model, which illustrates that the observed deficiencies
of RBM generated configurations persist in absence of conserved quantities.

In this subsection we use Monte Carlo (MC) generated configurations with conserved magnetization $M_0$ to train the RBM and compare the properties of RBM generated configurations
with those obtained from MC. We present results for both $M_0 = 0$ and $M_0 > 0$. The demanding task of generating 
configurations via RBM only allows to investigate rather small system sizes. The results presented in the following have been obtained with $L=8$ and $L=12$.

Fig. \ref{fig5} compares the RBM and MC averages for the commonly investigated energy and magnetization densities as a function of temperature $T$. For the energy density, see Fig. \ref{fig5}a,
we find the same good agreement between RBM and MC averages as observed for the non-conserved case in Ref. \cite{Torlai16}. Whereas the average energy density does not
hint at any major issues with the RBM generated configurations, this is different for the average absolute value of the magnetization density.
As shown in Fig. \ref{fig5}b,
the constraint, that the magnetization is $M_0=0$ in all MC configurations used to train RBM, 
is not captured by the machine learning algorithm, and, consequently, in many RBM generated
configurations $M \ne 0$ as there is an excess of one of the spin types. This yields an average absolute value of the magnetization that differs from that of the training configurations.
The shortcoming of RBM to identify and reproduce a conserved quantity is also expected to be encountered in other systems as well as for other conserved quantities.

\begin{figure}
 \centering \includegraphics[width=0.60\columnwidth,clip=true]{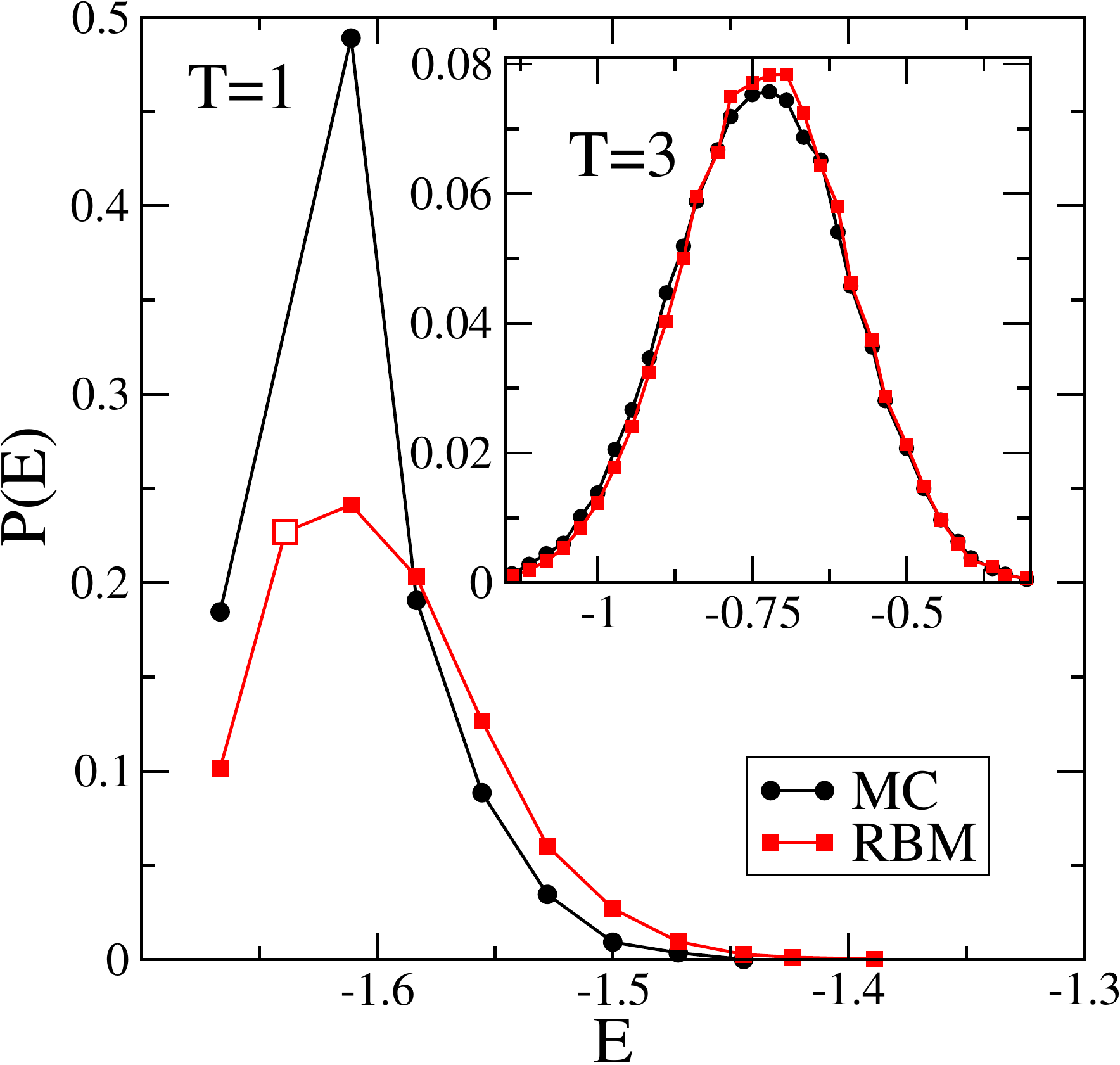}
 \caption{Probability distributions for the energy density for RBM and MC with $L=12$. The main image compares the probability distributions at $T=1$, i.e. below
the critical temperature, whereas the inset shows the
probability distribution at $T=3$, above the critical temperature. The white square indicates an energy value
that can not be accessed in MC simulations of the conserved-magnetization Ising model with $M_0=0$, but is present in almost 23\% of the the RBM generated configurations.
For $T=1$ we used 10,000 configurations to produce the probability distribution, whereas 100,000 generated configurations were used for $T=3$.
}
\label{fig6}
\end{figure}

\begin{figure}
 \centering \includegraphics[width=0.60\columnwidth,clip=true]{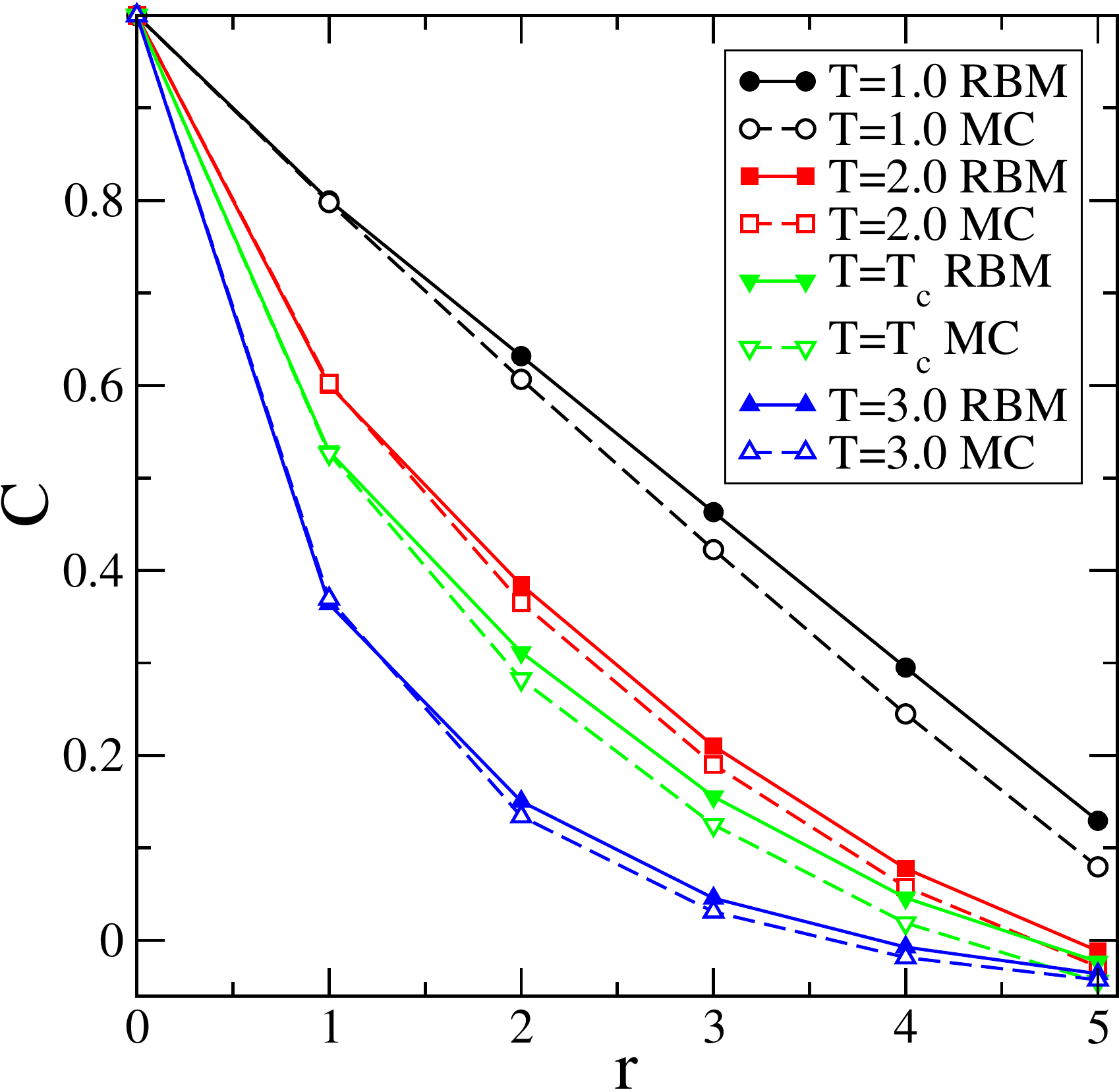}
 \caption{Comparison of the average space-dependent correlations $C(r)$ obtained from MC configurations with $M_0=0$ and from RBM configurations generated after
training with the same MC configurations. Data for four different temperatures are shown for a system with $L=12$. Error bars are smaller than the symbol sizes.
}
\label{fig7}
\end{figure}

%
The probability distribution for the energy density, displayed in Fig. \ref{fig6}, shows marked differences between RBM and MC, and this even though the average
energy densities closely agree, see Fig. \ref{fig5}a. Below $T_c$, and as exemplified by the data for $T=1$ in Fig. \ref{fig6}, RBM fails to learn the correct
weights for the different energy levels. Furthermore, as RBM does not recognize the conserved magnetization, it generates configurations with spin 
arrangements that break the conserved quantity and, in some cases, yield configurations with energies that are not accessible in the Ising model with
conserved order parameter. An example is provided in the figure by the large white rectangle. Differences in weights can also be seen for $T > T_c$, see the inset
for an example at $T=3$, especially in the increasing part at low energies and around the peak maximum.
We checked that these differences do not change substantially when doubling
the number of configurations used to train the RBM.

As the space-dependent correlation function $C(r)$ is proportional to the energy density for $r=1$ (indeed $C(r=1)=- \frac{1}{2} \varepsilon$, 
see equations (\ref{E}) and (\ref{C})), 
we have for $C(r=1)$ the same agreement between MC and RBM as we have for the average energy density. However, as shown in Fig. \ref{fig7},
marked differences show up for $r \ge 2$. 
These deviations between the spatial correlations computed from RBM and MC configurations 
point to challenges encountered by RBM when extracting information from the training configuration that go beyond the simple nearest neighbor
correlations. These deviations, which are largest for temperatures $T \le T_c$, are decreasing for larger temperatures and are expected to vanish
completely for very high temperatures where the typical configuration is a highly disordered one with only very small connected clusters and very short-ranged correlations.

\begin{figure}
 \centering \includegraphics[width=0.60\columnwidth,clip=true]{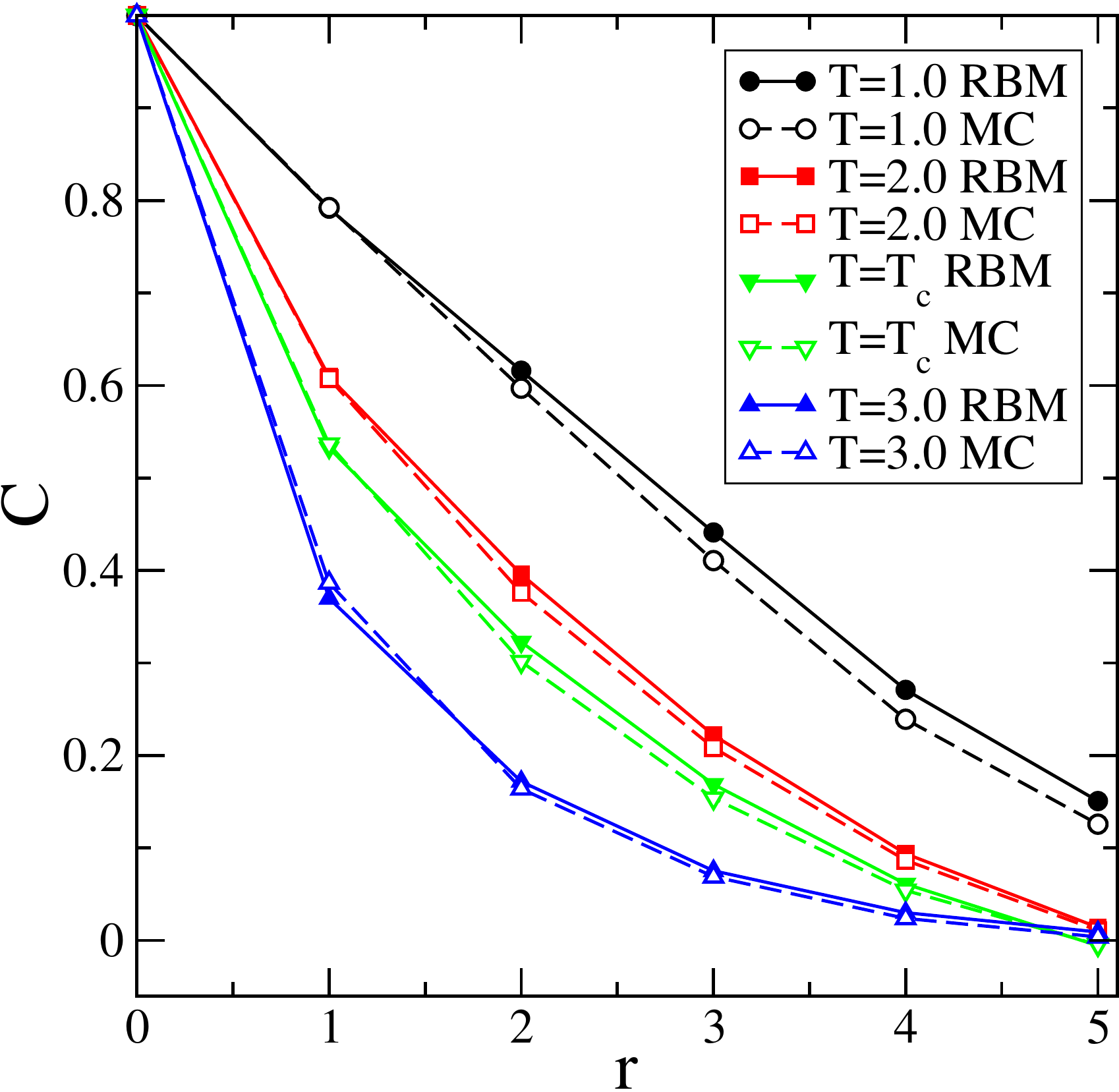}
 \caption{Comparison of the average space-dependent correlations $C(r)$ obtained from MC configurations with an excess of 30 spins 
of one of the spin types and from RBM configurations generated after
training with the same MC configurations. Data for four different temperatures are shown for a system with $L=12$. Error bars are smaller than the symbol sizes.
}
\label{fig8}
\end{figure}

The struggle of RBM to create a reasonable set of configurations for a fixed value of the magnetization is not restricted to the special case $M_0=0$,
where the same number of spins is encountered for each spin type, but it also persists for other values of $M_0$. For this we run Monte Carlo simulations with
spin exchanges for systems with fixed, but different, numbers of spins for both types, so that the magnetization is fixed
at some value $M_0 > 0$. The configurations obtained from these simulations are then
used to compute the MC quantities and to train a RBM in order to generate new configurations. These newly generated configurations
are then used to compute RBM quantities. In Fig. \ref{fig8} we show
as an example the difference in spatial correlations for configurations with $L=12$ when in the MC dataset 
the number of majority spins exceeds by 30 the number of minority spins, yielding
the constant value $M_0 = 0.2083$ for the magnetization. Systematic differences between RBM and MC are also observed for the other quantities investigated
in this study. RBM fares slightly better when increasing the value of the fixed magnetization $M_0$ as it is easier for
RBM to learn the energy distribution due to the fact that the number of accessible energies changes
with $M_0$, going from the largest number for $M_0=0$ to a single accessible energy for $M_0=1$.

\subsection{Ising systems with non-conserved magnetization}
 
\begin{figure}
 \centering \includegraphics[width=0.60\columnwidth,clip=true]{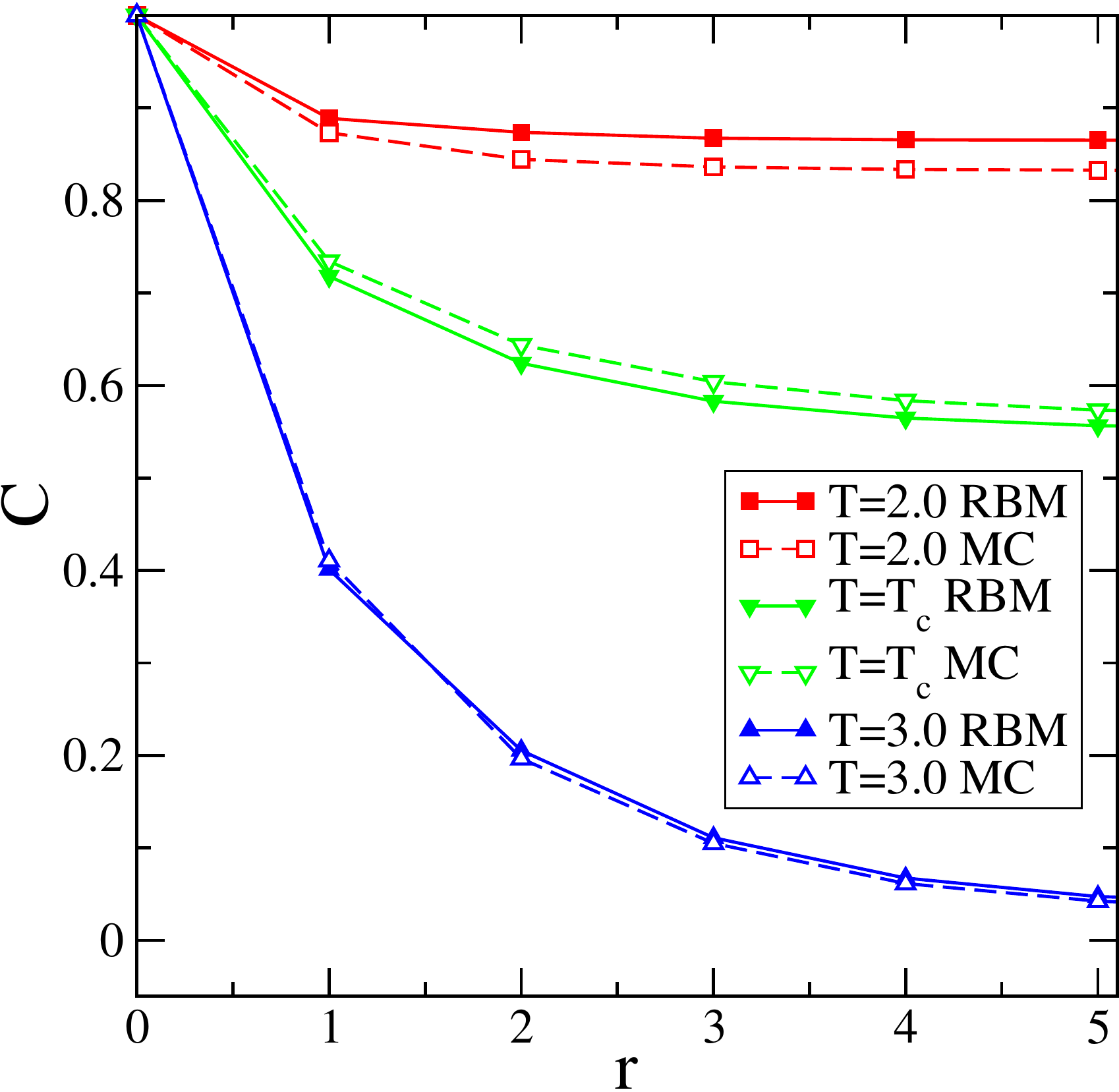}
 \caption{Comparison of the average space-dependent correlations $C(r)$ obtained from configurations generated in
Monte Carlo simulations of the non-conserved Ising model as well as 
from RBM configurations generated after
training with the same MC configurations. Data at different temperatures are shown for a system with $L=12$. Error bars are smaller than the symbol sizes.
}
\label{fig9}
\end{figure}

As the earlier accounts highlighting successes of RBM in creating configurations of classical systems \cite{Torlai16,Rao18} were dealing with
systems without conserved quantities, we found it of interest to have a fresh look at the Ising model without conserved magnetization,
following closely what has been done previously, but to compute other quantities than merely the average magnetization and energy. We did
check, though, that our results for the magnetization and energy densities agree with those obtained in \cite{Torlai16}

As shown in Fig. \ref{fig9}, even when trained using configurations without conserved magnetization, RBM fails to correctly capture
longer range correlations at temperatures $T \le T_c$. Similarly, deviations between RBM and MC are observed in the energy density
probability distributions (not shown). These are sobering observations, as they mean that RBM generated configurations
generically do not reproduce important physical aspects of the system used to train the machine (see \cite{Yevick20} for a related discussion
of shortcomings in the joint magnetization and energy probability distribution in small Ising systems computed from RBM generated configurations).
The fact that RBM generated configurations
do not have the same statistical properties as the configurations obtained from importance sampling Monte Carlo simulations  
makes it doubtful that machine learning generated configurations can be used safely for speeding up
numerical simulations, as proposed in \cite{Huang17,Liu17}.

\section{Conclusion}

Recent years have seen a strong increase of physicists' interest in supervised and unsupervised machine learning. Some of the most
promising applications are found in quantum physics \cite{Carleo17,Chng17,Broecker17,Torlai18} and in materials discovery \cite{Curtarolo03,Morgan05,Fischer06}. 
Classical systems have mainly been used as test cases in 
order to gain a better understanding of the power of machine learning algorithms by comparing their outputs with well understood results
from statistical physics. 

Previous studies of standard interacting spin systems have revealed that machine learning algorithms can identify different phases,
locate phase transition points, determine the order of a phase transition, and generate configurations that yield average values for 
magnetization and energy densities in agreement with values obtained from importance sampling Monte Carlo simulations. While these are remarkable achievements,
they are often qualitative or are dealing with a few selected averaged quantities that do not fully characterize the system.

The vast majority of machine learning studies of classical spin systems considered situations without conserved quantities. However, conserved quantities
restrict the accessible part of configuration space which often results for finite systems in modified physical properties. Two studies briefly
discussed the two-dimensional Ising model with conserved (zero) magnetization and showed that standard machine learning techniques (principal
component analysis \cite{Wang16} and support vector machine \cite{Ponte17}) allow to locate the phase transition point using the machine learning output. No past attempts were
made to use machine learning for more demanding tasks than the transition point location in systems with a conserved quantity.

In this paper we have presented two different types of results for the two-dimensional Ising model with conserved magnetization. In the first part of the paper
we have shown that supervised machine learning using convolutional neural networks not only allows to locate the phase transition temperature
with high precision, it also yields as output a quantity that behaves like an order parameter and displays a critical finite-size scaling
governed by the exponent of the two-dimensional Ising model. In the previous studies of the conserved-magnetization
Ising model the outputs from PCA and SVM merely allowed to locate the 
phase transition point without permitting to determine the value of a critical exponent through finite-size scaling. The second part of our paper has
revealed severe shortcomings of the configurations generated from a restricted Boltzmann machine trained with configurations obtained from the
conserved-magnetization Ising model. Not only does RBM not recognize a conserved quantity and therefore generates configurations with magnetizations
that differ from that of the training dataset, yielding energies forbidden in the original system, it also yields 
changes in weights for allowed energies as well as space-dependent correlations that differ from those obtained in importance sampling Monte Carlo simulations
of the original system. These observations have triggered us to also revisit the non-conserved Ising model: whereas RBM yields
for that system an acceptable agreement with Monte Carlo simulations for quantities like the average magnetization and the average energy \cite{Torlai16}, 
a closer look reveals that RBM does not yield very 
good approximations for the energy density probability distributions nor does it produce space-dependent correlations that agree with those
obtained from Monte Carlo simulations. Therefore, the statistical properties of configurations generated by RBM present marked differences from those used in
the training dataset, and this independently whether or not a conserved quantity is present.

Our results for the RBM configurations beg the question whether RBM generated configurations should be used at all. This will depend on the
application and on whether rough estimates of average quantities are needed or whether high precision data are required that fulfill the statistical
properties of the original physical system. Especially worrisome seem to be the proposals to use machine learning generated configurations
to speed up numerical simulations \cite{Huang17,Liu17}, as this approach may inject into the Markov chain configurations with statistical properties that differ from those
of the original system. It will depend on the system and on the physical question at hand whether this is acceptable and allows to obtain results that are
physically meaningful for the original system.

Our investigation exclusively dealt with classical spin systems, however similar issues with statistical properties can also be expected to
show up in some quantum applications. We hope that our work will trigger more rigorous approaches to use machine learning outputs in
the different fields of physics.

\begin{acknowledgments}
This work is in part supported by the US National Science Foundation through
grant DMR-1606814.
\end{acknowledgments}

\end{document}